\documentclass{ws-ijmpa}

\begin{document}

\markboth{J.A. Belinch\'on} {Bianchi I with variable $G$ and
$\Lambda$. Self-Similar approach.}

%
\catchline{}{}{}{}{}
%

\title{Bianchi I with variable $G$ and $\Lambda$. Self-Similar approach.}

\author{Jos\'{e} Antonio Belinch\'{o}n}

\address{Dept. Physics, ETS Arquitectura, UPM, Av. Juan de Herrera 4\\
Madrid, 28040, Espa\~{n}a\\
first\ abelcal@ciccp.es}

\maketitle

\begin{history}
\received{Day Month Year} \revised{Day Month Year} \comby{Managing
Editor}
\end{history}

\begin{abstract}
In this paper we study how to attack under the self-similarity
hypothesis a perfect fluid Bianchi I model with variable $G,$and
$\Lambda,$ but under the condition $\operatorname{div}T\neq0.$ We
arrive to the conclusion that: $G$ and $\Lambda$ are decreasing
time functions (the sing of $\Lambda$ depends on the equation of
state), while the exponents of the scale factor must satisfy the
conditions $\sum_{i=1}^{3}\alpha_{i}=1$ and
$\sum_{i=1}^{3}\alpha_{i}^{2}<1,$ $\forall\omega\in\left(
-1,1\right)  ,$ relaxing in this way the Kasner conditions. We
also show the connection between the behavior of $G$ and the Weyl
tensor.
\end{abstract}


\section{Introduction.}

Ever since Dirac first considered the possibility of a $G$ variable (see
\cite{1}), there have been numerous modifications of general relativity to
allow for a variable $G,$ nevertheless these theories have not gained wide
acceptance. However, recently (see \cite{2}-\cite{14}) a modification has
been proposed treating $G$ and $\Lambda$ as non-constants coupling scalars.
So it is considered $G$ and $\Lambda$ as coupling scalars within the
Einstein equations, $R_{ij}-\frac{1}{2}g_{ij}=GT_{ij}-\Lambda g_{ij},$ \
while the other symbols have their usual meaning and hence the principle of
equivalence then demands that only $g_{ij}$ and not $G$ and $\Lambda$ must
enter the equation of motion of particles and photons. In this way the usual
conservation law, $divT=0,$ holds. Taking the divergence of the Einstein
equations and using the Bianchi identities we obtain the an equation that
controls the variation of $G$ and $\Lambda.$ These are the modified field
equations that allow to take into account a variable $G$ and $\Lambda.$
Nevertheless this approach has some drawbacks, for example, it cannot
derived from a Hamiltonian, although there are several advantages in the
approach.

There are many publications devoted to study the variation of $G$ and $%
\Lambda$ in the framework of flat FRW symmetries (see for example \cite{2}-%
\cite{14}) and all this works have been extended to more complicated
geometries, like for example Bianchi I models, which represent the simplest
generalization of the flat FRW models (see for example \cite{Bes1}-\cite{SPS}%
. in the context of perfect fluids and \cite{Arbab}-\cite{Saha1} in the
context of viscous fluids). Bianchi I models are important in the study of
anisotropies.

But in our opinion, the problem arises when one try to solve the resulting
field equations (FE). It seems that it is unavoidable to make simplifying
hypotheses, or to impose ad hoc some particular behavior for some of the
quantities of the model, in order to obtain a exact solution to the FE.\
Such simplifying hypothesis are made for mathematical reason (in order to
reduce the number of unknowns) although are justified form the physical
point of view. Usually such assumptions or simplifying hypotheses follow a
power law, for example, the quantity $X$ follows a power law \ i.e. $%
X=X_{0}t^{\alpha},$ where $X_{0}$ is an appropriate dimensional constant, $\
t$ is the cosmic time (for example) and $\alpha\in\mathbb{R}$ (usually $%
\alpha\in\mathbb{Q},$ but this other question), and depending on the nature
of the quantity $X,$ $\alpha$ will be positive or negative. Actually we
think that although all these simplifying hypotheses are correct or at least
bring us to obtain correct results, it is not necessary to do that, since
they may be deduced from symmetry principles in such a way that one may
justify (deduce) them from a correct mathematical principle, and usually all
these approaches have physical meaning.

Therefore the main goal of this paper is to apply the well known tactic
(approach) of self similarity (SS) in order to study and find exact
solutions for a perfect fluid Bianchi I models with variable $G$ and $%
\Lambda,$ but under the condition $\operatorname{div}T\neq0,$ and
trying to make the lowest number of assumptions or neither. We
will try to show that with this approach all the usual simplifying
hypotheses may be deduced from a correct mathematical principle.

The paper is divided in the following sections. Section two is devoted to
outline all the ingredients as well as the field equations. In section
three, we study the model under the self-similarity hypothesis. We start
this section introducing briefly some ideas about self-similarity and
self-similar spacetimes. Once we have found the homothetic vector field we
go next to calculate the scale factors, where they obviously follow a power
law solution, as well as the derived quantities from them as the Hubble
parameter $H,$ the deceleration parameter $q$ and the shear $\sigma,$ since
they only depend of the scale factors and will be the same for all the
studied cases. We study four models. The first of them is the classical one
i.e. where $G$ is a true constant and the cosmological constant $\Lambda$
vanish. We have preferred starting with this model in other to check how
works the employed tactic. We emphasize that the obtained solution (that
will be the same in all the studied cases) satisfies the condition, $%
\sum_{i=1}^{3}\alpha_{i}=1$ and $\sum _{i=1}^{3}\alpha_{i}^{2}<1,$ relaxing
in this way the Kasner conditions ($\sum_{i=1}^{3}\alpha_{i}=$ $%
\sum_{i=1}^{3}\alpha_{i}^{2}=1,$ only valid for a vacuum solution), but it
is only valid for $\omega=1,$ i.e. ultra-stiff matter, and where $\left(
\alpha_{i}\right) _{i=1}^{3}$ are the exponents of the scale factors. If $%
\omega\neq1,$ the model collapse to the standard flat FRW one. We then study
the curvature invariants as well as the Weyl tensor and its invariant, and
end calculating the gravitational entropy. We show that the definitions of
gravitational entropy do not work well in this kind of spacetimes, the
self-similar ones, since these definitions are dimensionless which means
that this quantity remains constant along the homothetic trajectories of any
self-similar spacetime. We would like to relate in same way the behavior of
the Weyl tensor with the behavior of a $G$ time-varying. So after showing
that the tactic works correctly, we pass to study a simple model where we
consider that only vary $G.$ In this case we get the same solution as in the
above model with the same restriction for the equation of state i.e. the
solution is only valid if $\omega=1.$ In the third of the studied model we
consider only a $\Lambda$ time-varying. As we have pointed out previously
the solution is the same with regard to the exponents of the sale factors
i.e. they must satisfy the conditions $\sum_{i=1}^{3}\alpha _{i}=1$ and $%
\sum_{i=1}^{3}\alpha_{i}^{2}<1,$ but in this case this solution is only
valid iff $\omega\in\left( -1,1\right) $ i.e. $\omega\neq1,$ since if $%
\omega=1$, $\Lambda=0.$ We find always that $\Lambda$ behaves as a negative
decreasing time function. The last of the studied cases considers that both
\textquotedblleft constants\textquotedblright, $G$ and $\Lambda$ are
time-varying. We find that $G$ is a decreasing time function (i.e. has a
similar behavior as the Weyl tensor) and $\Lambda\thickapprox t^{-2}$, with $%
\Lambda<0,$ and as in the above case this solution is only valid $%
\forall\omega\in\left( -1,1\right) .$ If $\omega>1$ then
$\Lambda>0.$  We also show how to regain the \textquotedblleft
classical\textquotedblright\ case where $\operatorname{div}T=0.$
We end we a brief conclusions.

\section{The Model\label{Model}.}

Throughout the paper $M$ will denote the usual smooth (connected, Hausdorff,
4-dimensional) spacetime manifold with smooth Lorentz metric $g$ of
signature $(-,+,+,+)$. Thus $M$ is paracompact. A comma, semi-colon and the
symbol $\mathcal{L}$ denote the usual partial, covariant and Lie derivative,
respectively, the covariant derivative being with respect to the Levi-Civita
connection on $M$ derived from $g$. The associated Ricci and stress-energy
tensors will be denoted in component form by $R_{ij}(\equiv R^{c}{}_{jcd})$
and $T_{ij}$ respectively. A diagonal Bianchi I space-time is a spatially
homogeneous space-time which admits an abelian group of isometries $G_{3}$,
acting on spacelike hypersurfaces, generated by the spacelike KVs $\mathbf{%
\xi}_{1}=\partial_{x},\mathbf{\xi}_{2}=\partial_{y},\mathbf{\xi}%
_{3}=\partial_{z}$. In synchronous co-ordinates the metric is:%
\begin{equation}
ds^{2}=-dt^{2}+A_{\mu}^{2}(t)(dx^{\mu})^{2}   \label{sx1.2}
\end{equation}
where the metric functions $A_{1}(t),A_{2}(t),A_{3}(t)$ are functions of the
time co-ordinate only (Greek indices take the space values $1,2,3$ and Latin
indices the space-time values $0,1,2,3$). In this paper we are interested
only in \emph{proper diagonal} Bianchi I space-times (which in the following
will be referred for convenience simply as Bianchi I\ space-times), hence
all metric functions are assumed to be different and the dimension of the
group of isometries acting on the spacelike hypersurfaces is three.
Therefore we consider the Bianchi type I metric as
\begin{equation}
ds^{2}=-c^{2}dt^{2}+X^{2}(t)dx^{2}+Y^{2}(t)dy^{2}+Z^{2}(t)dz^{2},
\label{eq1}
\end{equation}
see for example (\cite{em}-\cite{Raycha}).

For a perfect fluid with energy-momentum tensor:
\begin{equation}
T_{ij}=\left( \rho+p\right) u_{i}u_{j}+pg_{ij},   \label{eq0}
\end{equation}
where we are assuming an equation of state $p=\omega\rho,\left(
\omega=const.\right) $. Note that here we have preferred to assume this
equation of state but as we will show in the following sections this
equation may be deduced from the symmetries principles as for example the
self-similar one. The $4-$velocity is defined as follows%
\begin{equation}
u=\left( \frac{1}{c},0,0,0\right) ,\qquad u_{i}u^{i}=-1.   \label{4-vel}
\end{equation}

The time derivatives of $G$ and $\Lambda$ are related by the Bianchi
identities
\begin{equation}
\left( R_{ij}-\frac{1}{2}Rg_{ij}\right) ^{;j}=\left( \frac{8\pi G}{c^{4}}%
T_{ij}-\Lambda g_{ij}\right) ^{;j},   \label{eq8}
\end{equation}
in this case this equation reads:%
\begin{equation}
\dot{\rho}+\rho\left( 1+\omega\right) \left( \frac{\dot{X}}{X}+\frac {\dot{Y}%
}{Y}+\frac{\dot{Z}}{Z}\right) =-\frac{\dot{\Lambda}c^{4}}{8\pi G}-\frac{\dot{%
G}}{G}\rho.   \label{laura3}
\end{equation}

Therefore the resulting field equations are:
\begin{align}
\frac{\dot{X}}{X}\frac{\dot{Y}}{Y}+\frac{\dot{X}}{X}\frac{\dot{Z}}{Z}+\frac{%
\dot{Z}}{Z}\frac{\dot{Y}}{Y} & =\frac{8\pi G}{c^{2}}\rho+\Lambda c^{2},
\label{eq3} \\
\frac{\ddot{Y}}{Y}+\frac{\ddot{Z}}{Z}+\frac{\dot{Z}}{Z}\frac{\dot{Y}}{Y} & =-%
\frac{8\pi G}{c^{2}}\omega\rho+\Lambda c^{2},  \label{eq4} \\
\frac{\ddot{X}}{X}+\frac{\ddot{Z}}{Z}+\frac{\dot{X}}{X}\frac{\dot{Z}}{Z} & =-%
\frac{8\pi G}{c^{2}}\omega\rho+\Lambda c^{2},  \label{eq5} \\
\frac{\ddot{X}}{X}+\frac{\ddot{Y}}{Y}+\frac{\dot{X}}{X}\frac{\dot{Y}}{Y} & =-%
\frac{8\pi G}{c^{2}}\omega\rho+\Lambda c^{2},  \label{eq6} \\
\dot{\rho}+\rho\left( 1+\omega\right) \left( \frac{\dot{X}}{X}+\frac {\dot{Y}%
}{Y}+\frac{\dot{Z}}{Z}\right) & =-\frac{\dot{\Lambda}c^{4}}{8\pi G}-\frac{%
\dot{G}}{G}\rho.   \label{eq9}
\end{align}

Now, we define
\begin{equation}
H=\left( \frac{\dot{X}}{X}+\frac{\dot{Y}}{Y}+\frac{\dot{Z}}{Z}\right) =3%
\frac{\dot{R}}{R}\text{ \ and \ \ }R^{3}=XYZ,\qquad q=\frac{d}{dt}\left(
\frac{1}{H}\right) -1,   \label{eq14}
\end{equation}

Since we have defined the 4-velocity by eq. (\ref{4-vel}) then the expansion
$\theta$ is defined as follows:
\begin{equation}
\theta:=u_{;i}^{i},\text{ \ \ \ \ \ \ \ \ }\theta=\frac{1}{c}\left( \frac{%
\dot{X}}{X}+\frac{\dot{Y}}{Y}+\frac{\dot{Z}}{Z}\right) =\frac{1}{c}H,
\end{equation}
and therefore the acceleration is: $a_{i}=u_{i;j}u^{j},$in this case $a=0,$
while the shear is defined as follows: $\sigma_{ij}=\frac{1}{2}\left(
u_{i;a}h_{j}^{a}+u_{j;a}h_{i}^{a}\right) -\frac{1}{3}\theta h_{ij},$%
\begin{equation}
\sigma^{2}=\frac{1}{2}\sigma_{ij}\sigma^{ij},\text{\ \ \ \ \ \ }\sigma ^{2}=%
\frac{1}{3c^{2}}\left( \left( \frac{\dot{X}}{X}\right) ^{2}+\left( \frac{%
\dot{Y}}{Y}\right) ^{2}+\left( \frac{\dot{Z}}{Z}\right) ^{2}-\frac{\dot{X}}{X%
}\frac{\dot{Y}}{Y}-\frac{\dot{X}}{X}\frac{\dot{Z}}{Z}-\frac{\dot{Y}}{Y}\frac{%
\dot{Z}}{Z}\right) ,   \label{defshear}
\end{equation}

\section{Self-similar solution.\label{SS}}

In general relativity, the term self-similarity can be used in two ways. One
is for the properties of spacetimes, the other is for the properties of
matter fields. These are not equivalent in general. The self-similarity in
general relativity was defined for the first time by Cahill and Taub (see
\cite{CT}, and for general reviews \cite{21}-\cite{Hall}). Self-similarity
is defined by the existence of a homothetic vector ${V}$ in the spacetime,
which satisfies
\begin{equation}
\mathcal{L}_{V}g_{ij}=2\alpha g_{ij},   \label{gss1}
\end{equation}
where $g_{ij}$ is the metric tensor, $\mathcal{L}_{V}$ denotes Lie
differentiation along ${V}$ and $\alpha$ is a constant. This is a special
type of conformal Killing vectors. This self-similarity is called homothety.
If $\alpha\neq0$, then it can be set to be unity by a constant rescaling of $%
{V}$. If $\alpha=0$, i.e. $\mathcal{L}_{V}g_{ij}=0$, then ${V}$ is a Killing
vector.

Homothety is a purely geometric property of spacetime so that the physical
quantity does not necessarily exhibit self-similarity such as $\mathcal{L}%
_{V}Z=dZ$, where $d$ is a constant and $Z$ is, for example, the pressure,
the energy density and so on. From equation (\ref{gss1}) it follows that
\begin{equation}
\mathcal{L}_{V}R^{i}\,_{jkl}=0,
\end{equation}
and hence
\begin{equation}
\mathcal{L}_{V}R_{ij}=0,\qquad\mathcal{L}_{V}G_{ij}=0.   \label{mattercoll}
\end{equation}
A vector field ${V}$ that satisfies the above equations is called a
curvature collineation, a Ricci collineation and a matter collineation,
respectively. It is noted that such equations do not necessarily mean that ${%
V}$ is a homothetic vector. We consider the Einstein equations
\begin{equation}
G_{ij}=8\pi GT_{ij},   \label{einstein}
\end{equation}
where $T_{ij}$ is the energy-momentum tensor.

If the spacetime is homothetic, the energy-momentum tensor of the matter
fields must satisfy
\begin{equation}
\mathcal{L}_{V}T_{ij}=0,   \label{emcoll}
\end{equation}
through equations~(\ref{einstein}) and (\ref{mattercoll}). For a perfect
fluid case, the energy-momentum tensor takes the form of eq. (\ref{eq0})
i.e. $T_{ij}=(p+\rho)u_{i}u_{j}+pg_{ij},$where $p$ and $\rho$ are the
pressure and the energy density, respectively. Then, equations~(\ref{gss1})
and (\ref{emcoll}) result in
\begin{equation}
\mathcal{L}_{V}u^{i}=-\alpha u^{i},\qquad\mathcal{L}_{V}\rho=-2\alpha
\rho,\qquad\mathcal{L}_{V}p=-2\alpha p.   \label{ssmu}
\end{equation}
As shown above, for a perfect fluid, the self-similarity of the spacetime
and that of the physical quantity coincide. However, this fact does not
necessarily hold for more general matter fields. Thus the self-similar
variables can be determined from dimensional considerations in the case of
homothety. Therefore, we can conclude homothety as the general relativistic
analogue of complete similarity.

From the constraints (\ref{ssmu}), we can show that if we consider the
barotropic equation of state, i.e., $p=f(\rho)$, then the equation of state
must have the form $p=\omega\rho$, where $\omega$ is a constant. This class
of equations of state contains a stiff fluid ($\omega=1$) as special cases,
whiting this theoretical framework. There are many papers devoted to study
Bianchi I models (in different context) assuming the hypothesis of
self-similarity (see for example \cite{HW}-\cite{griego}) but here, we would
like to try to show how taking into account this class of hypothesis one is
able to find exact solutions to the field equations within the framework of
the time varying constants.

The homothetic equations are given by eq. (\ref{gss1}) so it is a
straightforward task to find the homothetic vector field, where in this case
is as follows:%
\begin{equation}
V=t\partial_{t}+\left( 1-t\frac{\dot{X}}{X}\right) x\partial_{x}+\left( 1-t%
\frac{\dot{Y}}{Y}\right) y\partial_{y}+\left( 1-t\frac{\dot{Z}}{Z}\right)
z\partial_{z},   \label{HO1}
\end{equation}

Therefore, we have obtained the following behavior for the scale factors:%
\begin{equation}
X=X_{0}t^{\alpha_{1}},\qquad Y=Y_{0}t^{\alpha_{2}},\qquad Z=Z_{0}t^{\alpha
_{3}},
\end{equation}
with $X_{0},Y_{0},Z_{0}$ are integrating constants and $\left( \alpha
_{i}\right) _{i=1}^{3}\in\mathbb{R}.$ In this way we find that%
\begin{equation}
H=\left( \sum_{i=1}^{3}\alpha_{i}\right) \frac{1}{t}=\frac{\alpha}{t},\qquad
q=\frac{d}{dt}\left( \frac{1}{H}\right) -1=\frac{1}{\alpha}-1,\qquad
\sigma^{2}=\frac{1}{3c^{2}}\left( \sum_{i}^{3}\alpha_{i}^{2}-\sum_{i\neq
j}^{3}\alpha_{i}\alpha_{j}\right) \frac{1}{t^{2}}.
\end{equation}
with $\left( \alpha_{1}+\alpha_{2}+\alpha_{3}\right) =\alpha.$

In this section we are going to study several Bianchi I models and we will
show how it is possible to find exact solutions to the field equations
(without the condition $divT=0)$ under the hypothesis of SS.

The time derivatives of $G$ and $\Lambda$ are related by the Bianchi
identities i.e. eq. (\ref{eq9}) that in this case collapses to the following
one:%
\begin{equation}
\dot{\rho}+\rho\left( 1+\omega\right) H=f(t)=-\frac{\dot{\Lambda}c^{4}}{8\pi
G}-\frac{\dot{G}}{G}\rho,   \label{B1}
\end{equation}
where $f(t)$ is a function that depends on time and controls the time
variation of the constant $G$ or/and $\Lambda.$ If $G=const.$ and $\Lambda$
vanish then $f(t)=0,$ so the model collapses to the standard one. This idea
was pointed out by Rastal (see \cite{Rastall}) and improved (in the
theoretical framework of time varying constants) by Harko and Mak (see \cite%
{harko}).

Therefore the resulting field equations are (\ref{eq3}-\ref{eq6}) together
to the new one
\begin{equation}
\dot{\rho}+\rho\left( 1+\omega\right) H=f(t),   \label{Beq9}
\end{equation}

\subsection{\textquotedblleft Constants\textquotedblright\ constants. The
classical model.}

In this case we consider $f(t)=0,$ so this means that $G=const.$ and $\Lambda
$ vanish and therefore we get that from eq. (\ref{Beq9}) that
\begin{equation}
\dot{\rho}+\rho\left( 1+\omega\right) H=0,\qquad\Rightarrow\qquad\rho
=\rho_{0}t^{-(\omega+1)\alpha}.
\end{equation}

From the field equations (\ref{eq3}) we get that%
\begin{equation}
\rho_{0}=\frac{Ac^{2}}{8\pi G},\qquad\alpha=\frac{2}{(\omega+1)},
\label{m0}
\end{equation}
where $A=\alpha_{1}\alpha_{2}+\alpha_{3}\alpha_{1}+\alpha_{2}\alpha_{3}.$

The shear has the following behavior, $\sigma^{2}\neq0$, as it is observed $%
\sigma\rightarrow0$ as $\left( \alpha_{i}\rightarrow\alpha_{j}\right) . $ As
in the previous sections, we may calculate the coefficients $\left(
\alpha_{i}\right) $ by solving the following system of equations:%
\begin{align}
\alpha_{2}\left( \alpha_{2}-1\right) +\alpha_{3}\left( \alpha_{3}-1\right)
+\alpha_{3}\alpha_{2} & =-A\omega,  \label{m1} \\
\alpha_{1}\left( \alpha_{1}-1\right) +\alpha_{3}\left( \alpha_{3}-1\right)
+\alpha_{3}\alpha_{1} & =-A\omega, \\
\alpha_{2}\left( \alpha_{2}-1\right) +\alpha_{1}\left( \alpha_{1}-1\right)
+\alpha_{1}\alpha_{2} & =-A\omega, \\
\alpha(\omega+1) & =2,   \label{m4}
\end{align}
where $A=\alpha_{1}\alpha_{2}+\alpha_{3}\alpha_{1}+\alpha_{2}\alpha_{3},$
and $\alpha=\alpha_{1}+\alpha_{2}+\alpha_{3}.$

So we have the following solutions for this system of equations:%
\begin{align}
\alpha_{1} & =1-\alpha_{2}-\alpha_{3},\qquad\omega=1,  \label{Bsolss1} \\
\alpha_{1} & =\alpha_{2}=\alpha_{3}=\frac{2}{3\left( \omega+1\right) },
\label{Bsolss2}
\end{align}
as it is observed only solution (\ref{Bsolss1}) is interesting for us. The
second solution is the usual FRW one, so it is not \ interesting for us (see
Einstein\&de Sitter (\cite{EdS}) for $\omega=0$, and Harrison (\cite%
{Harrison}) $\forall\omega)$. Nevertheless we have found that solution (\ref%
{Bsolss1}) verifies the conditions
\begin{equation}
\alpha=\sum\alpha_{i}=1,\qquad\sum\alpha_{i}^{2}<1,
\end{equation}
but iff $\omega=1,$ (see \cite{HW}) while other authors claim that must be
satisfies the condition $\sum\alpha_{i}^{2}=1,$ (see \cite{Kasner}, \cite{SH}
and \cite{Jacobs}) and in particular, in this context (see \cite{griego}).
Nevertheless we have found that this solution only verifies the first of the
condition of the Kasner like solutions i.e. $\alpha=\sum\alpha_{i}=1,$ while
the second condition $\sum\alpha_{i}^{2}=1,$ it is not verified (see \cite%
{Kasner} \ and \ \cite{SH}). In this case we find that it is verified the
condition $\sum\alpha_{i}^{2}<1.$ Therefore we have found the same behavior
as the obtained one in (\cite{HW}). Before ending we would like to make a
little comment about the Kasner like solutions. If a solution of (\ref{m1}-%
\ref{m4}) verifies the relationships $\sum_{i}^{3}\alpha_{i}^{2}=%
\sum_{i}^{3}\alpha _{i}=1,$ i.e. they are Kasner's type (see \cite{Kasner},
\cite{SH} and in particular \cite{griego}), then this means that $%
A=\alpha_{1}\alpha_{2}+\alpha_{3}\alpha_{1}+\alpha_{2}\alpha_{3}=0$ (i.e.
the model is Ricci flat)$,$ which brings us to get the following result: $%
\alpha_{1}=\frac{1}{2}\left( 1-\alpha_{3}-\sqrt{1+2\alpha_{3}-3\alpha_{3}^{2}%
}\right) <0,$ $\alpha_{2}=\frac{1}{2}\left( 1-\alpha_{3}+\sqrt{%
1+2\alpha_{3}-3\alpha _{3}^{2}}\right) >0,$ $\forall\alpha_{3}\in\left(
0,1\right) ,$ we think that this class of solutions are unphysical and have
a pathological curvature behavior as it is shown bellow. Furthermore, as we
can see, if $A=0,$ then from eq. (\ref{m0}) we get $\rho=0,$ as it is
expected for this class of solutions (vacuum solutions) so they are not
interested for us. Nevertheless relaxing the condition $\sum\alpha_{i}^{2}=1,
$ to our result i.e. $\sum \alpha_{i}^{2}<1$, we are able to obtain
solutions whit $\left( \alpha _{i}\right) >0,\forall i$, and $\rho\neq0.$

Therefore we have obtained the following behavior for the main quantities:%
\begin{equation}
H=\frac{1}{t},\qquad\Longrightarrow\qquad q=0,
\end{equation}
so it is quite difficult to reconcile this model with the observational
data. With regard to the energy density we find that%
\begin{equation}
\rho=\frac{Ac^{2}}{8\pi G}t^{-2},\qquad\sigma^{2}=\frac{1}{3c^{2}}\left(
1+3A\right) \frac{1}{t^{2}},   \label{results1}
\end{equation}
and with regard to the constants $\left( \alpha_{i}\right) _{i=1}^{3}$ we
have that only obtain a BI solution iff $\ \alpha_{1}=1-\alpha_{2}-\alpha
_{3},$ (where furthermore we suppose that $\alpha_{2}\neq\alpha_{3})$ and
that this result only is possible if the equation of state is $\omega=1,$
i.e. ultra-stiff matter (see \cite{HW}). For a review of Bianchi I solutions
see for example (\cite{Harvey}).

With regard to the curvature behavior, we may see that the full contraction
of the Riemann tensor (see for example \cite{Caminati}-\cite{Barrow}) $%
I_{1}:=R_{ijkl}R^{ijkl},$ and the full contraction of the Ricci tensor, $%
I_{2}:=R_{ij}R^{ij},$ are:
\begin{equation}
I_{1}=\frac{K}{c^{4}t^{4}},\qquad I_{2}=\frac{4(-\alpha_{2}-\alpha_{3}+%
\alpha_{2}^{2}+\alpha_{2}\alpha_{3}+\alpha_{3}^{2})^{2}}{c^{4}t^{4}},
\end{equation}
where $K=K(\alpha_{i})=const\neq0,$ i.e.%
\begin{equation}
K=\left[ 3\left( \alpha_{2}^{2}+\alpha_{3}^{2}\right)
+2\alpha_{2}\alpha_{3}+9\alpha_{2}^{2}\alpha_{3}^{2}+3\left(
\alpha_{2}^{4}+\alpha _{3}^{4}\right) -6\left(
\alpha_{2}^{3}+\alpha_{3}^{3}\right) +\alpha _{2}\alpha_{3}\left( 6\left(
\alpha_{2}^{2}+\alpha_{3}^{2}\right) -8\left( \alpha_{2}+\alpha_{3}\right)
\right) \right] .
\end{equation}
and the scalar curvature $R$ is: $R=R_{i}^{i}.$

The non-zero components of the Weyl tensor are:%
\begin{align}
C_{1212} & =K_{1}t^{-2(1-\alpha_{1})},\qquad C_{1313}=K_{2}t^{-2(1-\alpha
_{2})},\qquad C_{1414}=K_{3}t^{-2(1-\alpha_{3})},  \nonumber \\
C_{2323} & =K_{4}t^{-2\alpha_{3}},\qquad
C_{2424}=K_{5}t^{-2\alpha_{2}},\qquad C_{3434}=K_{6}t^{-2\alpha_{1}},
\end{align}
where the numerical constants $\left( K_{i}\right) _{i=1}^{6}=K(\alpha
_{i})=const\neq0.$ As we can see with the obtained solution for $\left(
\alpha_{i}\right) _{i=1}^{3}$ the $Weyl\rightarrow\infty$ as $t\rightarrow0,$
in the next models we shall show that $G(t)$ has the same behavior as the
Weyl tensor. In a forthcoming paper we study models with $Weyl\rightarrow0$
as $t\rightarrow0$ (i.e. models that verify the Weyl tensor hypothesis) and
with a growing $G,$ i.e. in same way exists a relationship between both
quantities.

Now taking into account a very famous result by Hall et al (see \cite{hrv})
we may check that $\mathcal{L}_{V}C_{jkl}^{i}=0,$ as it is shown in (\cite%
{hrv}) if a vector field $V\in\mathfrak{X}(M),$ verifies the conditions $%
\mathcal{L}_{V}C_{jkl}^{i}=0,$ and $\mathcal{L}_{V}T_{ij}=0$ (as it is
known, if $V$ is a homothetic vector field, then it is also a matter
collineation)$,$ then $\mathcal{L}_{V}g=2g$ i.e. it is a homothetic vector
field, but in this case we have arrived to the conclusion that $\mathcal{L}%
_{V}g=2g\Longleftrightarrow\mathcal{L}_{V}T_{ij}=0$, and that it is also
verified the relationship $\mathcal{L}_{V}C_{jkl}^{i}=0.$

The Weyl scalar is defined as: $I_{3}:=C^{abcd}C_{abcd},$ as it is observed $%
I_{3},$ is also defined as follows: $I_{3}=I_{1}-2I_{2}+\frac{1}{3}R^{2},$
this definition is only valid when $n=4.$ Therefore, $I_{3}$ has the
following behavior%
\begin{equation}
I_{3}=\frac{\hat{K}}{c^{4}t^{4}},
\end{equation}
with $\hat{K}$ given by $\hat{K}=\frac{16}{3}\left[ \alpha_{2}^{2}+\alpha
_{3}^{2}-\alpha_{2}\alpha_{3}+3\alpha_{2}^{2}\alpha_{3}^{2}+\alpha_{2}^{4}+%
\alpha_{3}^{4}-2\left( \alpha_{2}^{3}+\alpha_{3}^{3}\right) +\alpha
_{2}\alpha_{3}\left( 2\left( \alpha_{2}^{2}+\alpha_{3}^{2}\right) -\left(
\alpha_{2}+\alpha_{3}\right) \right) \right] .$

The non-zero components of the electric part of the Weyl tensor are:%
\begin{equation}
E_{22}=\tilde{K}_{1}t^{-2(1-\alpha_{1})},\qquad E_{33}=\tilde{K}%
_{2}t^{-2(1-\alpha_{2})},\qquad E_{44}=\tilde{K}_{3}t^{-2(1-\alpha_{3})},
\end{equation}
while the magnetic part of the Weyl tensor vanish,\textbf{\ }$H_{ij}=0.$

The gravitational entropy is defined as follows (see \cite{gron1}-\cite%
{gron2}):%
\begin{equation}
P^{2}=\frac{I_{3}}{I_{2}}=\frac{I_{1}-2I_{2}-\frac{1}{3}R^{2}}{I_{2}}=\frac{%
I_{1}}{I_{2}}+\frac{1}{3}\frac{R^{2}}{I_{2}}-2.   \label{penrose}
\end{equation}
finding that
\begin{equation}
P^{2}=const.\neq0,
\end{equation}
note that $P^{2}=I_{3}/I_{2}.$ As have been pointed out by Nicos Pelavas et
al (see \cite{Lake}) this definition is not an acceptable candidate for
gravitational entropy along the homothetic trajectories of any self-similar
spacetime. Nor indeed is any \textquotedblleft
dimensionless\textquotedblright\ scalar. This implies that $I_{3}/I_{2}$ is
constant along timelike homothetic trajectories. As a consequence, (\ref%
{penrose}) does not provide a measure of gravitational entropy along
homotheticities and therefore $I_{3}/I_{2}$ cannot be a candidate for a
measure of gravitational entropy in self-similar spacetimes.

\subsection{$G-$variable.}

In this case we are going to consider that only vary \textquotedblleft
constant\textquotedblright\ $G.$ This only possible if we take into account
the condition $divT\neq0$ and therefore $f(t)=-\frac{G^{\prime}}{G}\rho,$ so
eq. (\ref{Beq9}) collapses to the following one.
\begin{equation}
\frac{\rho^{\prime}}{\rho}+\frac{G^{\prime}}{G}=-\left( 1+\omega\right)
\frac{\alpha}{t},\qquad\Longrightarrow\qquad\rho G=t^{-\left( 1+\omega
\right) \alpha},   \label{Blau1}
\end{equation}

From the field equations (\ref{eq3}) we get that%
\begin{equation}
G\rho=\frac{c^{2}}{4\pi}\frac{A}{\alpha\left( 1+\omega\right) }\frac {1}{%
t^{2}},\qquad\alpha=\frac{2}{(\omega+1)}.
\end{equation}

The shear has the following behavior, $\sigma^{2}\neq0$, as it is observed $%
\sigma\rightarrow0$ as $\left( \alpha_{i}\rightarrow\alpha_{j}\right) . $ As
in the previous sections, to calculate the coefficients $\left( \alpha
_{i}\right) $ we need to solve the filed equations obtaining the same system
of equations as in the above case i.e. eqs. (\ref{m1}-\ref{m4}), so we get
the same solution as in the above case i.e. $\sum_{i=1}^{3}\alpha_{i}=1,$
and $\sum_{i=1}^{3}\alpha_{i}^{2}<1,$ and only valid if $\omega=1. $

Therefore we have obtained the following behavior for the main quantities:%
\begin{equation}
H=\frac{1}{t},\qquad\Longrightarrow\qquad q=0,
\end{equation}
and
\begin{equation}
G\rho=\frac{Ac^{2}}{8\pi}t^{-2},\qquad\sigma^{2}=\frac{1}{3c^{2}}\left(
1-3A\right) \frac{1}{t^{2}},
\end{equation}
note that this result is quite similar to the obtained one in the last
solution i.e. the obtained one in eq. (\ref{results1}), but we are not able
to get a separate behavior for the quantities $G$ and $\rho$.

\subsection{$\Lambda-$variable.}

In this case we consider only the variation of the cosmological constant $%
\Lambda,$ so eq. (\ref{Beq9}) yields
\begin{equation}
\dot{\rho}+\rho\left( 1+\omega\right) \left( \frac{\dot{X}}{X}+\frac {\dot{Y}%
}{Y}+\frac{\dot{Z}}{Z}\right) =-\frac{\dot{\Lambda}c^{4}}{8\pi G},
\label{laugL}
\end{equation}
and therefore from the field equations (\ref{eq3}) we get that%
\begin{equation}
\Lambda^{\prime}=-\frac{A}{c^{2}}\frac{2}{t^{3}}-\frac{8\pi G}{c^{4}}%
\rho^{\prime},   \label{dorota1}
\end{equation}
and hence%
\begin{equation}
\rho=\frac{c^{2}}{4\pi G}\frac{A}{\left( 1+\omega\right) \alpha}\frac {1}{%
t^{2}}.
\end{equation}

Now, we next to calculate the quantity $\Lambda,$ from eq. (\ref{dorota1})
we get%
\begin{equation}
\Lambda=\frac{A}{c^{2}}\left( 1-\frac{2}{\left( 1+\omega\right) \alpha }%
\right) \frac{1}{t^{2}},
\end{equation}
in this way it is observed that%
\begin{equation}
\Lambda=\Lambda_{0}t^{-2},\qquad\left\{
\begin{array}{c}
\Lambda_{0}>0\Longleftrightarrow(\omega+1)\alpha>2 \\
\Lambda_{0}=0\Longleftrightarrow(\omega+1)\alpha=2 \\
\Lambda_{0}<0\Longleftrightarrow(\omega+1)\alpha<2%
\end{array}
\right. .
\end{equation}

The shear has the following behavior, $\sigma^{2}\neq0$, by hypothesis. As
in the previous sections, we calculate the coefficients $\left( \alpha
_{i}\right) $ from following system of equations:%
\begin{align}
\alpha_{2}\left( \alpha_{2}-1\right) +\alpha_{3}\left( \alpha_{3}-1\right)
+\alpha_{3}\alpha_{2} & =A\left( \frac{\alpha-2}{\alpha}\right) ,
\label{lau1} \\
\alpha_{1}\left( \alpha_{1}-1\right) +\alpha_{3}\left( \alpha_{3}-1\right)
+\alpha_{3}\alpha_{1} & =A\left( \frac{\alpha-2}{\alpha}\right) , \\
\alpha_{2}\left( \alpha_{2}-1\right) +\alpha_{1}\left( \alpha_{1}-1\right)
+\alpha_{1}\alpha_{2} & =A\left( \frac{\alpha-2}{\alpha}\right) ,
\label{lau2}
\end{align}
where $A=\alpha_{1}\alpha_{2}+\alpha_{3}\alpha_{1}+\alpha_{2}\alpha_{3},$
and $\alpha=\alpha_{1}+\alpha_{2}+\alpha_{3}.$

So we have the following solutions for this system of equations:%
\begin{align}
\alpha_{1} & =\alpha_{2}=\alpha_{3},  \label{Bsolsys1} \\
\alpha_{1} & =1-\alpha_{2}-\alpha_{3},   \label{Bsolsys3}
\end{align}
as it is observed solution (\ref{Bsolsys1}) is not interesting for us, since
it is unphysical (in this context). Only the second solution has physical
meaning and it is valid $\forall\omega\in\left( -1,1\right] $. Therefore, we
have found that the solution is $\sum_{i=1}^{3}\alpha_{i}=1,$ and $\sum
_{i=1}^{3}\alpha_{i}^{2}<1,$ but in occasion this solution is only valid if $%
\forall\omega\in\left( -1,1\right] .$

Therefore we have obtained the following behavior for the main quantities:%
\begin{equation}
H=\frac{1}{t},\qquad\Longrightarrow\qquad q=0,
\end{equation}
while with regard to the energy density we find that%
\begin{equation}
\rho=\frac{c^{2}}{4\pi G}\frac{A}{\left( 1+\omega\right) }\frac{1}{t^{2}},%
\text{ }
\end{equation}
\ so, if $\omega<-1\Longrightarrow\rho$ is negative (phantom cosmologies),
for the rest of the values of $\omega,$ i.e. $\omega\in(-1,1],$ $\rho$ is a
decreasing function on time.

The cosmological \textquotedblleft constant\textquotedblright\ behaves as
follows
\begin{equation}
\Lambda=\Lambda_{0}t^{-2},\qquad\Lambda_{0}=\left\{
\begin{array}{l}
\Lambda_{0}=0\Longleftrightarrow\omega=1, \\
\Lambda_{0}<0,\qquad\forall\text{ }\omega\in(-1,1)%
\end{array}
\right. ,
\end{equation}
so we have found that $\Lambda$ is a \textquotedblleft\emph{negative
decreasing function}\textquotedblright\ on time. Note that $\Lambda_{0}>0$
iff $\omega>1.$ As we can see this solution is quite different of the
previous ones, since here we have obtained a solution type Bianchi I $\forall
$ $\omega\in(-1,1)$ while in the previous ones this only happens if $%
\omega=1.$ Here if $\omega=1$ then we regain the first of the studied cases
i.e. which one where $\Lambda$ vanish and $G$ behaves as a true constant.

\subsection{$G\&\Lambda-$variable.}

In this case we are going to consider that both \textquotedblleft
constants\textquotedblright\ $G$ and $\Lambda$ vary, therefore eq. (\ref%
{Beq9}) yields

\begin{equation}
\dot{\rho}+\rho\left( 1+\omega\right) H=-\frac{\dot{\Lambda}c^{4}}{8\pi G}-%
\frac{\dot{G}}{G}\rho,   \label{dorota2}
\end{equation}
so from the field equations (\ref{eq3}) and (\ref{eq9}) \ and following the
same steps as in the above models we get that
\begin{equation}
G\rho=\frac{c^{2}}{4\pi}\frac{A}{(\omega+1)\alpha}t^{-2}.
\end{equation}
as we can see it is verified the relationship $G\rho\thickapprox t^{-2},$ as
it is expected. In fact it is impossible to separate both functions (to get
the behavior of both functions independently), to do that we need to impose
a condition, but precisely we are trying to avoid such way.

Now taking into account again eq. (\ref{eq3}) we get%
\begin{equation}
\Lambda=\Lambda_{0}t^{-2},\qquad\Lambda_{0}=\frac{A}{c^{2}}\left( 1-\frac {2%
}{(\omega+1)\alpha}\right) ,
\end{equation}
in this way it is observed that%
\begin{equation}
\Lambda=\Lambda_{0}t^{-2},\qquad\left\{
\begin{array}{c}
\Lambda_{0}>0\Longleftrightarrow(\omega+1)\alpha>2 \\
\Lambda_{0}=0\Longleftrightarrow(\omega+1)\alpha=2 \\
\Lambda_{0}<0\Longleftrightarrow(\omega+1)\alpha<2%
\end{array}
\right. .
\end{equation}

The shear behaves (see eq. (\ref{defshear})) as follows: $\sigma^{2}\neq0,$
by hypothesis. In order to find the value of constants $\left( \alpha
_{i}\right) ,$ we make that them verify the field eqs. so in this case we
get the following system of eqs.:%
\begin{align}
\alpha_{2}\left( \alpha_{2}-1\right) +\alpha_{3}\left( \alpha_{3}-1\right)
+\alpha_{3}\alpha_{2} & =A\left( \frac{\alpha-2}{\alpha}\right) , \\
\alpha_{1}\left( \alpha_{1}-1\right) +\alpha_{3}\left( \alpha_{3}-1\right)
+\alpha_{3}\alpha_{1} & =A\left( \frac{\alpha-2}{\alpha}\right) , \\
\alpha_{2}\left( \alpha_{2}-1\right) +\alpha_{1}\left( \alpha_{1}-1\right)
+\alpha_{1}\alpha_{2} & =A\left( \frac{\alpha-2}{\alpha}\right) ,
\end{align}
(note that this system is the same as eqs. (\ref{lau1}-\ref{lau2}) where $%
A=\alpha_{1}\alpha_{2}+\alpha_{3}\alpha_{1}+\alpha_{2}\alpha_{3},$ and $%
\alpha=\alpha_{1}+\alpha_{2}+\alpha_{3}.$ Therefore we obtain the same
solution as in the last studied case, i.e. $\sum_{i=1}^{3}\alpha_{i}=1,$ and
$\sum_{i=1}^{3}\alpha_{i}^{2}<1,$ with $\omega\in\left( -1,1\right] ,$
finding in this way the following behavior for the main quantities:%
\begin{equation}
H=\frac{1}{t},\qquad\Longrightarrow\qquad q=0,
\end{equation}
and \ with regard to the product $G\rho$ we get
\begin{equation}
G\rho=\frac{c^{2}}{4\pi}\frac{A}{(\omega+1)}t^{-2},
\end{equation}
but we cannot say anything more. The cosmological \textquotedblleft
constant" behaves as follows
\begin{equation}
\Lambda=\Lambda_{0}t^{-2},\qquad\Lambda_{0}=\left\{
\begin{array}{l}
\Lambda_{0}=0\Longleftrightarrow\omega=1 \\
\Lambda_{0}<0,\qquad\forall\text{ }\omega\in(-1,1)%
\end{array}
\right. ,
\end{equation}
so we have found that $\Lambda$ is a negative decreasing function on time.
As in the above case we get a positive cosmological \textquotedblleft
constant\textquotedblright\ if $\omega>1.$

In order to try to find a separate behavior for the functions $\rho$ and $G,
$ we may suppose that
\begin{equation}
\rho=\rho_{0}t^{-a},\qquad G=G_{0}t^{a-2},\qquad\Longrightarrow\qquad G\rho=%
\frac{c^{2}}{4\pi}\frac{A}{(\omega+1)}t^{-2}=Kt^{-2},
\end{equation}
with $a\in\mathbb{R}^{+},$ i.e. for example we may choice
\begin{equation}
G=\frac{c^{2}}{4\pi\rho_{0}}\frac{A}{(\omega+1)}t^{a-2}=\frac{K}{\rho_{0}}%
t^{a-2},
\end{equation}
therefore, it is verified the field eq. (\ref{dorota2}) for all the possible
values of $a.$ We may find other possibilities as for example
\begin{equation}
\rho=\rho_{0}t^{-2+a},\qquad G=G_{0}t^{-a}
\end{equation}
with $a\in\left( -\infty,2\right) .$ At this point we would like to stress
the relationship between the behavior of the Weyl tensor and the behavior of
$G(t),$ since $Weyl\longrightarrow\infty$ as $t\rightarrow0$ in the same way
as $G(t).$

But if $a=(\omega+1)$ then we regain the condition
$\operatorname{div}T=0$ as well as $f(t)=0,$ i.e.
\begin{equation}
\operatorname{div}T=\dot{\rho}+\rho\left( 1+\omega\right) H=0=-\frac {\dot{\Lambda}%
c^{4}}{8\pi G}-\frac{\dot{G}}{G}\rho=f(t).
\end{equation}
but this case has been already studied in (\cite{tony}).

\section{Conclusions.\label{Conclu}}

We have shown how to attack a perfect fluid Bianchi I with $G$ and
$\Lambda$ variable under the condition $\operatorname{div}T\neq0,$
taking account only the hypothesis of SS. Our arguments exploit
the symmetry properties of homothetically self-similar spacetimes.
These calculations are of physical interest since self-similar
spacetimes are very widely studied \cite{CC} and are, for example,
believed to play an important role in describing the asymptotic
properties of more general models. In this way we have shown that
it is not necessary to make any assumption
\textquotedblleft\emph{ad hoc\textquotedblright} or to take into
account any previous hypothesis or considering any hypothetical
behavior for any quantity since all these hypotheses could be
deduced from the symmetry principles, as for example the
self-similar hypothesis SS (for other approaches see \cite{tony}).
As have seen to get the solution under the SS hypothesis it very
simple since once we have calculate the homothetic vector field it
is a trivial task to obtain the behavior of the scale factors,
where the must follow a power law
solution, $X=X_{0}t^{\alpha_{1}},$ $Y=Y_{0}t^{\alpha_{2}},$ $%
Y=Y_{0}t^{\alpha_{3}},$ in this way we only need to calculate the value of
theses exponents i.e. $\left( \alpha_{i}\right) _{i=1}^{3},$ since the rest
of the quantities will depend on these values. We have started studying the
simplest case where $G$ is a true constant and $\Lambda$ vanish, i.e. the
classical model in order to check how works the purposed method. For this
model we arrive to the already known result $\sum\alpha_{i}=1,$ and $%
\sum\alpha_{i}^{2}<1,$ stressing that this result is only valid if the
equation of state verifies the relationship $\omega=1,$ i.e. the result is
only valid for ultra-stiff matter, otherwise the model collapses to the FRW
solution. We have discussed why it is not possible to get the vacuum $\left(
\rho=0\right) $ Kasner solution, $\sum\alpha_{i}=1=\sum\alpha_{i}^{2}=1.$ We
think that this class of solutions are unphysical since necessarily one of
the scale factors must be a decreasing time function (maybe such class of
solutions would have any interest in the study of singularities). At the
same time, we have studied the curvature invariants, $I_{1}$ and $I_{2}$,
i.e. the Kretschmann's scalars, showing that the obtained solution is
singular as well as the Weyl tensor and its scalar, $I_{3}$. We have
performed all these calculations in order to show that there is a
relationship between the behavior of the Weyl tensor and the behavior of the
variable Newton constant $G(t).$ All these considerations are valid for the
rest of the studied models. We also have calculated the gravitational
entropy, $P^{2}=I_{3}/I_{2},$ showing that this definition is not valid for
self-similar spacetimes since this quantity is dimensionless and therefore
remains constant. Actually we have arrived to the same conclusion as the
obtained one by Nicos Pelavas et al (see \cite{Lake}). Furthermore if one
gets the Kasner solution then $I_{2}=0,$ i.e. the model collapses to a Ricci
flat model and therefore $P^{2}=\infty.$

With regard to the second of the studied models, where only vary $G(t),$ we
have shown that it is not possible to get a separate behavior of the
\textquotedblleft constants\textquotedblright\ $G$ and $\rho,$ obtaining $%
G\rho\sim t^{-2},$ as it is expected. For this model we have obtained the
same solution as in the previous case i.e. that the exponents of the scale
factors must satisfy the relationships $\sum\alpha_{i}=1,$ and $\sum\alpha
_{i}^{2}<1,$ and only valid for $\omega=1.$ As we have shown in the last of
the studied models in order to get a separate behavior between $G$ and $\rho,
$ it is possible to follow several ways but in this paper we are trying to
do the slightest number of hypotheses and in any case to avoid to make
previous assumptions on the behavior of any quantity.

In the third of the studied models, where only vary the
cosmological constant $\Lambda,$ we have found that $\Lambda$ is a
\emph{negative decreasing function} on time
$\forall\omega\in\left( -1,1\right) $ and a positive time
decreasing function if $\omega>1$. If $\omega=1$ then $\Lambda$
vanish, so the exponents verify the same
relationship as above but in this case this solution is only valid if $%
\omega\in\left( -1,1\right) $ since if $\omega=1$ then we get the first of
the studied models.

In the fourth model, we have considered that both constant vary.
In this case we arrive to similar conclusions as in the above
cases, i.e. $\Lambda$ is a \emph{negative decreasing function} on
time $\forall\omega\in\left( -1,1\right) $, vanish if $\omega=1$
and it is a positive decreasing time function if $\omega>1$ while
$G\rho\sim t^{-2}$ and the exponents of the scale factors must
satisfy the relationships $\sum\alpha _{i}=1,$ and
$\sum\alpha_{i}^{2}<1,$ valid $\forall\omega\in\left( -1,1\right)
.$ In this occasion we have made an assumption on the behavior of
$G$ in of to try to know its behavior finding in this way that is
a decreasing time function on time. It is quite surprising result
since in similar models with FRW symmetries this quantity is
always growing. At the same time we have shown the similitude
between its behavior and the behavior of the Weyl tensor. Both
quantities tend to infinite as $t$ runs to zero We have finished
howing how to regain, in a trivial way, the condition
$\operatorname{div}T=0.$

\end{document}